\documentclass [12pt]{article}
\usepackage{graphicx,amssymb,amsfonts,latexsym,amsmath,amsthm,times}
\usepackage{epsfig}
\usepackage{fancyhdr}
\usepackage{color}
\usepackage[english]{babel}
\numberwithin{equation}{section}

\newtheorem{theorem}{Theorem}

\newtheorem{conjecture}[theorem]{Conjecture}

\begin{document}


\thispagestyle{plain}

\vspace*{2cm} \normalsize \centerline{\Large \bf About Bifurcational Parametric Simplification}

\vspace*{1cm}

\centerline{\bf V.Gol'dshtein, N.Krapivnik$^{a}$, G.Yablonsky$^{b}$ \footnote{Corresponding
author. E-mail: vladimir@bgu.ac.il}}

\vspace*{0.5cm}

\centerline{$^{a}$ Department of Mathematics, Ben Gurion University of the Negev,
P.O.B. 653, 84105 Beer-Sheva, Israel}

\centerline{$^{b}$ Parks College of Engineering, Activation and Technology, Saint Louis University} \centerline {Lindell Blvd, 3450, St. Louis MO 63103, USA }


\vspace*{1cm}

\noindent {\bf Abstract.}
A concept of ``critical" simplification
was proposed by Yablonsky and Lazman in 1996 \cite{Critical} for
the oxidation of carbon monoxide over a platinum catalyst using a
Langmuir-Hinshelwood mechanism. The main observation was a simplification
of the mechanism at ignition and extinction points. The critical simplification
is an example of a much more general phenomenon that we call  \emph{a bifurcational
parametric simplification}. Ignition and extinction points are points
of equilibrium multiplicity bifurcations, i.e., they are points of a corresponding
bifurcation set for parameters.

Any bifurcation produces a dependence between system parameters. 
This is a mathematical explanation and/or justification of the ``parametric
simplification". It leads us to a conjecture that ``maximal
bifurcational parametric simplification" corresponds
to the ``maximal bifurcation complexity."

This conjecture can have practical applications for experimental study,
because at points of ``maximal bifurcation complexity'' the number of
independent system parameters is minimal and all other parameters
can be evaluated analytically or numerically.

We illustrate this method by the case of the simplest possible bifurcation,
that is a multiplicity bifurcation of equilibrium and we apply this
analysis to the Langmuir mechanism. Our analytical study
is based on a coordinate-free version of the method of invariant manifolds
(proposed recently \cite{BGG2006}). As a result we obtain a more
accurate description of the ``critical (parametric) simplifications."

With the help of the ``bifurcational parametric simplification"
kinetic mechanisms and reaction rate parameters may be readily identified
from a monoparametric experiment (reaction rate vs. reaction parameter).

\vspace*{0.5cm}




\vspace*{1cm}

\setcounter{equation}{0}
\section{Introduction and methodology}
By a standard linguistic definition, a bifurcation is a place where something
divides into two branches. Typical examples are: bifurcation of aorta
is the region in which the abdominal aorta bifurcates into the left
and right branches; river bifurcation, the forking of a river
into its distributaries; bifurcation of trachea, etc... For all these
examples,  bifurcations are places where the main stream is divided
into two or more streams. This dynamical interpretation corresponds 
to geometric changes of objects.

This simplistic meaning of  bifurcation can be extended. Roughly
speaking, a bifurcation is a region of qualitative changes in the
system's dynamical behavior. It can be induced by changes in geometry,
but also by other causes.

The notion of ``bifurcation'' in mathematics was introduced by Henri
Poincar{\'e} \cite{Poin} at the end of nineteen century for changes in the multuplicity 
of equalibrium systems of ordinary differential equations
(ODE). Nowadays,  this notion is used for systems of partial differential
equations (PDE) and all other types of continuous and discrete mathematical
models. Any mathematical model, except for a set of variables that describes
its dynamical behavior, also has a set of parameters that are ``permanent''
comparatively with variables. In reality these parameters are control
parameters and the system's asymptotic dynamics (for example, the equilibrium
multiplicity) depends on its values.

Roughly speaking, a mathematical bifurcation is a qualitative change
of asymptotic dynamics due to a smooth and slow change in the values of the parameters.
A typical example is the famous Hopf bifurcation that describes a change
from an equilibrium to oscillations. These qualitative changes can
be continuous and slow (bifurcation theory) but can be also fast and
even very fast (catastrophe theory).

Let us recall that ``bifurcation theory studies and classifies
phenomena characterized by sudden shifts in behavior arising from
small changes in circumstances, analyzing how the qualitative nature
of equation solutions depends on the parameters that appear in the
equation" (\cite{Arnold}). Analysis of complex bifurcation
points is a classical subject of the theory of dynamical system (\cite{Arnold}).

Mathematical Catastrophe theory was invented by Rene Thom in the book
\emph{Stabilite Structurelle et Morphogenese, }published in 1972 \cite{Thom}.
He proposed a classification of simple catastrophes using geometric
properties of surfaces in multidimensional spaces. By E.C. Zeeman
``the world is full of sudden transformations and unpredictable divergences''
that are subjects in Catastrophe theory \cite{Zeem}. One of the possible explanations
of such ``sudden transformations'' is the existence of fast sub-processes
into an original system, that relax very quickly to its new asymptotic
state. This fast relaxation is very hard to observe.

A set of parameters where a bifurcation or a catastrophe occurs is
called a bifurcation set $B$. It is a subset of the space of parameters
$P$ and the dimension of the bifurcation set is less than the dimension of
$P$. Typically it is a surface in $P$.  The dimension of $P$ ($\dim P$)
is the number of independent parameters. Because $\dim B$ is less then
$\dim P$, the number of independent parameters of $B$ is less than $\dim P$,
and a corresponding part of the parameters $m=\dim P-\dim B$ depends
on others.

\emph{ We call this phenomenon as ``bifurcational parametric simplification''.}

The number $m$ is called \emph{a bifurcational simplification number.}

\begin{conjecture}
The maximal bifurcational simplification number corresponds to a bifurcation
of ``maximal complexity."
\end{conjecture}

 The conjecture is more philosophical than it looks. The maximal bifurcational simplification 
 number is clear, but what
is \emph{a bifurcation of maximal complexity?} This question arises because  there 
 exist different types of bifurcations: change of stability,
Hopf bifurcation, multiplicity of limit cycles, non local bifurcations etc... In our opinion, a
more accurate understanding of \emph{the maximal bifurcation complexity
}depends on modeling phenomena.

The bifurcational parametric simplification was first described in
papers of Yablonsky et al. \cite{Critical}, \cite{SimpleModel} under
the name of ``critical simplification'' for the famous Langmuir
absorption model of heterogeneous catalytic kinetics. They observed
an essential simplification of the kinetics at extinction and ignition
points that permitted them to find simple connections between the  values of
``bifurcation parameters" and kinetic constants
\cite{SimpleModel}, \cite{YablBook}. As ``bifurcation
parameters" were used the maximal kinetic rate $R_3$, 
is an observable quantity. At ignition and extinction points $R_3$
is equal to the kinetic constants of absorption and desorption, that in
turn can be employed to applications involving kinetic constant measurements.

The Langmuir absorption model combine fast and slow kinetic
processes. Ignition and extinction points are points of fast bifurcations
(catastrophes).

\subsection {Discussion about "maximal complexity"}

We start from an historical remark.  In 1944, Lev Landau noticed that near the loss of stability the amplitude of the emergent "principal motion" satisfies a very simple equation \cite{Landau1944}.  It is an example of the "bifurcational parametric simplification".

In the case of an isothermic detailed kinetics a corresponding bifurcation set for steady states multiplicity  is an algebraic variety. By Whitney theorem \cite{Whitney1965} any algebraic variety admits a Whitney stratification. It is natural to conjecture that the stratum of minimal dimension corresponds to the "maximal complexity". Perhaps numerical algorithms for the Whitney stratification will be useful for an evaluation of the "maximal complexity". It seems to us that such algorithms have to be combined with a subdivision to fast and slow sub-processes. 

In the applied bifurcation theory there are constructed some interesting algorithms for numerical evaluation of the stratum of minimal dimension (maximal co-dimension), see \cite{Kuznetsov2004} and \cite{Khibnik1987} that can be useful to an evaluation of the "maximal complexity" and description of many possible transition trajectories in the vicinity of critical points. It was done for isothermal chemical system \cite{Khibnik1993}. Of course, such analysis of sophisticated near-steady-state behavior combined with a separation of fast and slow sub-processes looks natural. However it still does not provide with a robust knowledge on relationships between model parameters which determine this critical behavior. 

\subsection{Bifurcational parametric simplification of chemical kinetic models}
We propose here an accurate mathematical description of the bifurcational
parametric simplification for kinetic models in the case of the simplest
possible bifurcation of equalibriums (steady state) multiplicity. Our justification
is based on the bifurcation theory of dynamic systems. We also use 'slow-fast' dynamics, i.e. a coordinate free version of the singular perturbation theory, as a technical tool for an accurate evaluation of asymptotic analytic expressions that combine different reaction parameters
\cite{BGG2006}.

The following questions play a key role in this study: 
\begin{itemize}
\item How do the characteristics of multiple steady states and its bifurcations
depend on the kinetic parameters? 
\item How can process properties be readily related to experimental results? 
\end{itemize}
Let us explain more formally the main ideas of the bifurcational parametric
simplification analysis in the case of homogeneous kinetic systems.

In the vector notation the system of governing equations of a homogeneous
kinetic system can be written as 
\[
\frac{d\psi}{dt}=F(\psi,K)
\]
 where $\psi=(\psi_{1},\psi_{2},...,\psi_{n})$ is the thermochemical
state vector and $K=(k_{1},k_{2},...,k_{s})$ is the kinetic parameter
vector. (A detailed description of such models can be found in the
next section.)

Suppose our model has multiple steady states $\psi_{1,s},\psi_{2,s},...,\psi_{l,s}$
that are all consistent solutions of the functional system 
\begin{equation}
0=F(\psi,K).\label{Main}
\end{equation}

These steady states $\psi_{1,s},\psi_{2,s},...,\psi_{l,s}$ are functions
of the kinetic parameter vector $K=(k_{1},k_{2},...,k_{s})$, i.e.
\[
\psi_{1,s}=g_{1}(K),\psi_{2,s}=g_{2}(K),...,\psi_{l,s}=g_{l}(K).
\]
 The simplest possible bifurcation is a coincidence of two steady states
$\psi_{1,s}=g_{1}(K)=g_{2}(K)=\psi_{2,s}$. These bifurcations can be termed as multiplicity bifurcations. The multiplicity bifurcation happens
if all coordinates of both steady states coincide, i.e. we obtain
$n$ functional/algebraic equations 
\begin{equation}
g_{1}(k_{1},k_{2},...,k_{s})=g_{2}(k_{1},k_{2},...,k_{s})\label{eq:kinetic}
\end{equation}
 between kinetic parameters $k_{1},k_{2},...,k_{s}$. Therefore there
exists at least one functional dependence between the kinetic parameters
and at most $n$ functional dependencies.  The number of independent kinetic
constants $I$ depends on the structure of the steady states system
\ref{Main}.

For this case $Q=n-I$ is the bifurcational simplification number.
Any additional bifurcation point produces an additional functional
equation for kinetic parameters.

The bifurcational parametric simplification with the maximal simplification
number $Q$ corresponds to a bifurcation of ``maximal
complexity." It is clear that in our case the maximum critical simplification is just a set of kinetic parameters. It is just a set of kinetic constants (a value $K_{0}=(k_{1,0},k_{2,0},...,k_{s,0})$)
for which there exists a maximal number of functional dependencies.
 (\ref{eq:kinetic}).

\subsection*{Realization of the bifurcational parametric simplification analysis
of kinetic systems}

Here we propose an efficient and quite simple algorithm adapted to models of chemical kinetics. Main ingredients of this algorithm are:

-	coordinate transformation that reverts the original model to the slow-fast one\cite{BGG2006};

-	the method of invariant manifolds for slow-fast dynamics\cite{GS1992};

-	bifurcation analysis of invariant manifolds\cite{GS1992}

We apply this algorithm to the Langmuir mechanism. We corrected some inaccuracies of previous studies by Yablonsky et al using a more delicate algorithm for revealing slow-fast dynamics \cite{BGG2006}, in particular we found that the point (0, 0) is not a steady-state.  The main invariant of slow invariant manifolds is described as a bifurcation of ``maximal complexity'' of the Langmuir model.

\subsection*{Possible practical applications}

A practical application is usually the weak point of any theoretical algorithm. We hope that  experimental studies of practical systems for bifurcational kinetic parameters
can help in  evaluating kinetic parameters by using
analytic expressions for observable quantities. These analytical expressions exist due to the bifurcational parametric simplifications phenomenon .

\subsection*{General Conclusions and discussion}
\begin{itemize}
\item Any simple bifurcation point produces at least one functional dependence
between kinetic parameters. 
\item A complex bifurcation point (for example, coincidence of more then two
steady states) produces a number of functional dependencies that correspond
in complexity. 
\item At a bifurcation point of ``maximal complexity''  the number of independent
parameters is minimal. This means that all other kinetic parameters
can be analytically recalculated. 
\item It is an open question as to how one can use bifurcation critical simplification
for experimental study, but at least for ignition an extinction points
it looks possible. 
\end{itemize}

\subsection*{Conclusions about the Langmuir model}
\begin{itemize}
\item All steady states, their type, stability and attraction domains are
classified. Results about type and stability of steady states mainly
confirm previous study \cite{Critical} but are more accurate. It
was proved that $(0,0)$ is not a steady state as was claimed in \cite{Critical}.
Attraction domains have not been previously studied.  
\item All bifurcation points which are points of the multiplicity change have been studied. Similarly to \cite{Critical}, ignition and extinction bifurcation points have been observed. Types of these points were analyzed using the special coordinate transformation that is a subject of the theory of singular perturbed vector fields \cite{BGG2006}.

\item Bifurcation parameters $B_{2}$ for ignition and $B_{1}$ for extinction
were introduced and analyzed. At bifurcation points, corresponding parameters are vanishing. Bifurcation points are analyzed in details.
Results are illustrated with the help of invariant slow manifolds (curves)
by a number of figures. 
\item  At the bifurcation point of ``maximal complexity'' both parameters are vanishing. This point provides with maximal critical simplification.
This type of bifurcation, which previously was not observed  or analyzed, plays a crucial
role in our study. 
\item The first approximation of slow invariant manifold is analyzed
mainly in a vicinity of the point $(0,0)$. This analysis permits us to qualify
the  bifurcation type of ``maximal complexity'' and to evaluate
reaction rates. 
\end{itemize}

\vspace*{1cm}

\setcounter{equation}{0}
\section{Langmuir mechanism}

It is well known that the simplest mechanism for interpreting isothermal
critical phenomena in heterogeneous catalysis is the Langmuir
mechanism. It is a typical non-linear three -step adsorption
mechanism. For $CO$ oxidation on platinum, this mechanism is written
as follows: 
\begin{enumerate}
\item $2Z+O_{2}\leftrightarrow2ZO$; 
\item $Z+CO\leftrightarrow ZCO$; 
\item $ZO+ZCO\to2Z+CO_{2}$. 
\end{enumerate}
Here $Z$ is the free active catalyst, $ZO$ and $ZCO$ are species
adsorbed on the catalyst surface. The first and second step are considered
to be reversible.

We will divide the analysis of the problem into two parts: linear
analysis (analysis of the stoichiometry matrix) and non-linear analysis
(asymptotic analysis of bifurcations). The dimension of a kernel of
the stoichiometric matrix give the relationship between rates of reaction;
from the bifurcation analysis we obtain an additional relationship
between reaction rates.

We study a more simple case of irreversible first reaction. It is valid for 
wide domain of concentrations and pressures \cite{YablBook}
\begin{enumerate}
\item $2Z+O_{2}\to2ZO$; 
\item $Z+CO\leftrightarrow ZCO$; 
\item $ZO+ZCO\to2Z+CO_{2}$. 
\end{enumerate}

\subsection{Kernel of the stochiometric matrix}

In this case the stoichiometric matrix for $x:=[ZO],y:=[ZCO],z:=[Z]$
is

\[
S=\left(\begin{array}{cccc}
2 & 0 & 0 & -1\\
0 & 1 & -1 & -1\\
-2 & -1 & 1 & 2
\end{array}\right)
\]
 Combining the stoichiometry matrix $S$ with the rate reaction
vector $(R_{1},R_{2},R_{-2},R_{3})$ gives us the following kinetic model
 in terms of kinetic rates 
\[
\frac{dx}{dt}=2R_{1}-R_{3}
\]
 
\[
\frac{dy}{dt}=R_{2}-R_{-2}-R_{3}
\]
 
\[
\frac{dz}{dt}=-2R_{1}-R_{2}+R_{-2}+2R_{3}
\]

Here $R_{1}=k_{1}P_{O_{2}}z^{2},\, R_{2}=k_{2}P_{CO}z,\, R_{-2}=k_{-2}y,\, R_{3}=k_{3}xy$.

By standard calculations the rank of the stoichiometry matrix $S$
equals two. It means that the third row of the matrix is a linear
combination on the first and second rows, $Row_{3}=-(Row_{1}+Row_{2})$.
This relation gives us the existence of a well known linear integral which relates to catalyst mass conservation low
\cite{SimpleModel}

\[
x+y+z=1.
\]
 Because rank $S$ is $2$,  the system has no other linear integrals.

Hence, the corresponding kinetic model of the open catalytic system
can be reduced to

\[
\frac{dx}{dt}=2k_{1}P_{O_{2}}(1-x-y)^{2}-k_{3}xy
\]
 
\[
\frac{dy}{dt}=k_{2}P_{CO}(1-x-y)-k_{-2}y-k_{3}xy.
\]

This system has an invariant domain $U=[0,1]\times[0,1]$ in the $x,y$-phase
plane.

The corresponding steady state model is:

\begin{equation}
2R_{1}=R_{3}\label{stochiometry1}
\end{equation}
 
\begin{equation}
R_{2}-R_{-2}=R_{3}.\label{stochiometry2}
\end{equation}

The partial pressures $P_{O_{2}}$ and $P_{CO}$ are considered to
be parameters of our model.

\textbf{Conclusion.} Equations (\ref{stochiometry1}) and (\ref{stochiometry2})
show that only two of four system parameters $k_{1}P_{O_{2}},k_{2}P_{CO},k_{-2},k_{3}$
are independent at any steady state.

We rewrite the steady state system (\ref{stochiometry1}, \ref{stochiometry2})
in $(x,y)$ phase coordinates

\[
2k_{1}P_{O_{2}}(1-x-y)^{2}-k_{3}xy=0
\]
 
\[
k_{2}P_{CO}(1-x-y)-k_{-2}y-k_{3}xy=0.
\]

This system has an obvious steady state solution $(1,0)$. By simple
calculations we obtain the following cubic equation for other possible
steady states: 
\[
-2k_{1}P_{O_{2}}k_{3}^{2}x^{3}+(2k_{1}P_{O_{2}}k_{3}^{2}-4k_{1}P_{O_{2}}k_{-2}k_{3}-k_{3}^{2}k_{2}P_{CO})x^{2}
\]
 
\[
+(4k_{1}P_{O_{2}}k_{-2}k_{3}-2k_{1}P_{O_{2}}k_{-2}^{2}-k_{3}k_{2}^{2}P_{CO}^{2}-k_{-2}k_{3}k_{2}P_{CO})x+2k_{1}P_{O_{2}}k_{-2}^{2}=0.
\]
This is a cubic equation for the variable $x$. Corresponding values of $y$ can be evaluated from the previous system.

In  typical catalytic processes the elementary reaction
between two species on the catalytic surface $x$ and $y$ ($ZO+ZCO\to2Z+CO_{2}$) is fast,
i.e. $k_{3}>>k_{1}P_{O_{2}},k_{2}P_{CO},\, k_{-2}$. This permits us
to introduce a small parameter $\varepsilon:=1/k_{3}$ and write the
discriminant of the previous cubic equation in the following form:
\[
(-2k_{1}P_{O_{2}}+k_{2}P_{CO})^{2}(k_{-2}^{2}-8k_{1}P_{O_{2}}k_{-2}+2k_{2}P_{CO}k_{-2}+k_{2}^{2}P_{CO}^{2})-
\]
 
\[
\varepsilon4k_{1}P_{O_{2}}(24k_{1}^{2}P_{O_{2}}^{2}k_{-2}^{2}+k_{2}P_{CO}(k_{2}P_{CO}+k_{-2})(2k_{2}^{2}P_{CO}^{2}+2k_{2}P_{CO}k_{-2}-k_{-2}^{2})-
\]
\[
2k_{1}P_{O_{2}}k_{-2}(10k_{2}P_{CO}^{2}+k_{2}P_{CO}k_{-2}+k_{-2}^{2}))
\]
 
\[
-\varepsilon^{2}4k_{1}P_{O_{2}}^{2}k_{2}^{2}P_{CO}^{2}k_{-2}^{2}(-k_{-2}^{2}+8k_{2}^{2}P_{CO}^{2}+8k_{-2}(3k_{1}P_{O_{2}}+k_{2}P_{CO}))
\]
 
\[
-\varepsilon^{3}32k_{1}^{3}P_{O_{2}}^{3}k_{2}^{2}P_{CO}^{2}k_{-2}^{4}.
\]

In the zero approximation $\varepsilon=0$,  the discriminant becomes
very simple 
\begin{equation}
(k_{2}P_{CO}-2k_{1}P_{O_{2}})^{2}(k_{-2}^{2}-8k_{1}P_{O_{2}}k_{-2}+2k_{2}P_{CO}k_{-2}+k_{2}^{2}P_{CO}^{2}).\label{discriminant}
\end{equation}
 It can be presented as a product of two terms: $k_{2}P_{CO}-2k_{1}P_{O_{2}}$
and $k_{-2}^{2}-8k_{1}P_{O_{2}}k_{-2}+2k_{2}P_{CO}k_{-2}+k_{2}^{2}P_{CO}^{2}$.
The corresponding quantities 
\begin{equation}
\begin{array}{c}
B_{1}:=1-\frac{k_{2}P_{CO}}{2k_{1}P_{O_{2}}};\\
\ B_{2}:=k_{-2}^{2}-8k_{1}P_{O_{2}}k_{-2}+2k_{2}P_{CO}k_{-2}+k_{2}^{2}P_{CO}^{2}=\\
=\left(k_{2}P_{CO}+k_{-2}\right)^{2}-8k_{1}P_{O_{2}}k_{-2},
\end{array}\label{bifparameters}
\end{equation}
 will be used later as bifurcation parameters. A bifurcation happens
if $B_{1}=0$, or $B_{2}=0$. If $B_{1}=B_{2}=0$, then the corresponding
bifurcation has maximal complexity.

\vspace*{1cm}
\setcounter{equation}{0}
\section{Transformation of the Langmuir system to a slow-fast system}

The fast reaction rate $k_{3}xy$ is presented in both equations. The
original system does not have any fast-slow division in the original coordinate system $(x,y)$.

By the main concept in the theory of singularly perturbed vector fields
\cite{BGG2006}, \cite{BGM2008}, it is possible to construct another orthogonal
coordinate system where the original model becomes a slow-fast one.
In the case of the Langmuir system it is only a simple rotation.

An obvious candidate to a small parameter is $\varepsilon:=1/k_{3}$.
We use the following orthogonal coordinates transformation 
\[
u=\frac{x-y}{\sqrt{2}},
\]

\[
v=\frac{x+y}{\sqrt{2}}.
\]

With respect to the new coordinates, the system has the form 
\[
\frac{du}{dt}=\sqrt{2}k_{1}P_{O_{2}}(1-\sqrt{2}v)^2-\frac{k_{2}}{\sqrt{2}}P_{CO}(1-\sqrt{2}v)+\frac{k_{-2}}{2}(v-u),
\]
 
\[
\varepsilon\frac{dv}{dt}=\frac{u^{2}-v^{2}}{\sqrt{2}}+\frac{\varepsilon}{2\sqrt{2}}\left(4k_{1}P_{O_{2}}(1-\sqrt{2}u)^{2}+2k_{2}P_{CO}(1-\sqrt{2}v)-\sqrt{2}k_{-2}(v-u)\right).
\]
This is a standard slow-fast system (singularly perturbed system)
that combines singular and regular perturbations in the second
equation. The singular perturbation induces the small parameter $\varepsilon$ before
$\frac{dv}{dt}$, and the regular perturbation induces  the same small parameter before the second term in the right hand side of the second equation.

\subsection{Zero approximation $\varepsilon=0$ for the singular perturbation}

At this stage, all the machinery in the standard singular perturbation
theory can be applied to the analysis of the system, including its
decomposition to slow and fast subsystems and reduction of the near
steady state dynamics to dynamics on invariant slow manifolds (slow
curves in our model). 

The zero approximation $\varepsilon=0$ ($k_{3}\rightarrow\infty$)
of slow invariant curve is given by equating $\varepsilon$ to zero
before $\frac{dv}{dt}$ at the second equation. The slow curves equation
is 
\begin{equation}
\frac{u^{2}-v^{2}}{\sqrt{2}}+\frac{\varepsilon}{2\sqrt{2}}\left(4k_{1}P_{O_{2}}(1-\sqrt{2}u)^{2}+2k_{2}P_{CO}(1-\sqrt{2}v)-\sqrt{2}k_{-2}(v-u)\right)=0.\label{eq:accurate}
\end{equation}

To simplify analysis of the steady states we put $\varepsilon=0$ for the regular perturbation 
in this equation. The corresponding approximation of slow curves is represented
by $u^{2}-v^{2}=0$. It has two branches $u+v=0$ and $u-v=0$, or
in the original coordinates $x=0$ and $y=0$. Both branches of the slow
invariant manifold are stable (attractive).

Let us remark that the two branches $u+v=0$ and $u-v=0$ have an intersection
point $(0,0)$ where the last approximation is problematic. Around
the point $(0,0)$ a more accurate approximation is necessary. It
is easy to check that $(0,0)$ is not a steady state of the original
system (see\ref{stochiometry1}, \ref{stochiometry2}). This problem
is discussed in detail in Appendix 2. 

In the original coordinates the branches are $x=0$ and $y=0$. Hence
any steady state at this approximation has zero of its coordinates,
and can not be used for the evaluation of the reaction rate $R_{3}=k_{3}xy$
at any steady state, because it gives the reaction rate zero. 
This means
that we have to use a more accurate approximation \ref{eq:accurate}
for $R_{3}=k_{3}xy$. We shall use the steady state relationships of reaction
rates (\ref{stochiometry1}, \ref{stochiometry2}) for evaluations of
$R_{3}$.

\subsection{Multiplicity of bifurcation parameters}

I) For the first branch $v=u$ or $y=0$ of the slow curve the multiplicity
bifurcation parameter is 
\[
B_{1}=1-\frac{k_{2}P_{CO}}{2k_{1}P_{O_{2}}}
\]
 (see \ref{bifparameters}).

There are only one or two steady states on the branch $v=u$ or $y=0$
in the original coordinate system. The first one is $(1,0)$ and does
not depends on $B_{1}$.

The second one is 
\[
\left(1-\frac{k_{2}P_{CO}}{2k_{1}P_{O_{2}}},0\right).
\]
 If $B_{1}>0$ this steady state is stable. If $B_{1}<0$ then the
second steady state does not belong to the invariant domain $U$, i.e
it has no  physical meaning. The bifurcation value of $B_{1}$ is
equal $0$, i.e in the bifurcation point 
\begin{equation}
k_{2}P_{CO}=2k_{1}P_{O_{2}}.\label{eq:b_1=00003D0}
\end{equation}

\textbf{Conclusion} 
At the bifurcation point $B_{1}=0$, the parameters $k_1P_{O_2}$ and $k_2P_{CO}$ are dependent. This dependence is a bifurcational simplification.  The change of sign $B_{1}$ from positive to negative corresponds
to extinction, that follows from the evaluation of the reaction rates (next
subsection).

II) For the second branch $v=-u$ or $x=0$ of the slow curve the
multiplicity bifurcation parameter is 
\[
B_{2}=\left(k_{-2}\right)^{2}+2k_{-2}\left(k_{2}P_{CO}-4k_{1}P_{O_{2}}\right)+\left(k_{2}P_{CO}\right)^{2}
\]
 (see \ref{bifparameters}).

Let us remark that parameters $B_{1}$ and $B_{2}$ are not completely
independent. By simple calculation we obtain that $B_{2}=\left(k_{-2}\right)^{2}+2k_{-2}\left(-4k_{1}P_{O_{2}}B_{1}-k_{2}P_{CO}\right)+\left(k_{2}P_{CO}\right)^{2}=\left(k_{2}P_{CO}-k_{-2}\right)^{2}-8k_{-2}k_{1}P_{O_{2}}B_{1}$.
From this relation we conclude that if $B_{1}<0$,then  $B_{2}>0$
and if $B_{1}=0$ we have $B_{2}=\left(k_{2}P_{CO}-k_{-2}\right)^{2}\geq0$,
and the situation $B_{1}<0,B_{2}<0$ is not possible.

There exist two, one or zero steady states on this branch, depending
on the sign of $B_{2}$. For $B_{2}<0$ steady states does not exist.
Existence of one steady states corresponds to $B_{2}=0$. For $B_{2}>0$
there exist two steady states on $x=0$. Let us remark that $k_{-2}$
is the leading kinetic parameter, because for $k_{-2}=0$ the bifurcation
parameter $B_{2}$ is always positive and the bifurcation does not
exists.

\textbf{Conclusion.} At the bifurcation point $B_{2}=0$ there exists a functional relation between parameters $k_1P_{O_2}$, $k_2P_{CO}$, $k_{-2}$. It is a bifurcational simplification as well. The change
in the sign of $B_{2}$ from positive to negative corresponds to ignition,
that follows from the evaluation of the reaction rates (next subsection).

\subsection{Calculation of reaction rates}

We calculate the reaction rates at the steady states on the first 
slow curve $y=0$ using the zero approximation for $R_{1}$ and $R_{2}$
and the steady state relations between rates $R_{3}=2R_{1}$ and $R_{3}=R_{2}-R_{-2}$
for the evaluations of $R_{3}$ and $R_{-2}$ .

\textit{We use the following standard representation of the steady
states coordinate $x_{s}$, $y_{s}$ based on the regular perturbation
theory: 
\[
(x_{s}=x_{0}+\varepsilon x_{1}+...,y_{s}=y_{0}+\varepsilon y_{1}+...).
\]
 Hence $R_{3}=\frac{1}{\varepsilon}((x_{0}+\varepsilon x_{1}+...)(y_{0}+\varepsilon y_{1}+...))$.
In zero approximation $\varepsilon=0$ we have $x_{0}=0$ or $y_{0}=0$.
Therefore $R_{3}=x_{1}y_{0}$ or $R_{3}=x_{0}y_{1}$.} 

\vspace{0.5cm}

\textit{The first branch of the slow curve is $y=0$.} 

\vspace{0.5cm}

At the steady state $(1,0)$ all reaction rates equal  zero. The
second steady state 
\[
\left(1-\frac{k_{2}P_{CO}}{2k_{1}P_{O_{2}}},0\right)
\]
 exists for $B_{1}>0$. At this steady state the reaction rates are 
\[
R_{1}=\frac{k_{2}^{2}P_{CO}^{2}}{4k_{1}P_{O_{2}}},
\]
 
\[
R_{2}=\frac{k_{2}^{2}P_{CO}^{2}}{2k_{1}P_{O_{2}}},
\]
 
\[
R_{-2}=0,
\]
 
\[
R_{3}=2R_{1}=\frac{k_{2}^{2}P_{CO}^{2}}{2k_{1}P_{O_{2}}}.
\]
Clearly, we have an equality
\[
2R_{1}=R_{2}-R_{-2}
\]
In this approximation the rate of the second reverse reaction at the first branch is zero. Hence all reactions of the detailed 
mechanism can be considered to be irreversible.

{\bf Remark.}
In the first regular approximation $(x_{0}+\epsilon x_{1})$, at the first branch the rate of $CO$ desorption (the second reverse reaction) is zero. But for more accurate approximations it is positive, but small enough. Our approximation is of the  order $O(\epsilon)$ for singular perturbations.

\vspace{0.5cm}

\textit{The second branch of the slow curve is $x=0$.} 

\vspace{0.5cm}

At the case $B_{2}>0$. There exist two steady states (on this branch):
\[
\left(0,1-\frac{k_{-2}+k_{2}P_{CO}\pm\sqrt{B_{2}}}{4k_{1}P_{O_{2}}}\right).
\]
 Corresponding reaction rates are:

\[
R_{1}=k_{1}P_{O_{2}}\left(\frac{k_{-2}+k_{2}P_{CO}\pm\sqrt{B_{2}}}{4k_{1}P_{O_{2}}}\right)^{2},
\]
 
\[
R_{2}=k_{2}P_{CO}\left(\frac{k_{-2}+k_{2}P_{CO}\pm\sqrt{B_{2}}}{4k_{1}P_{O_{2}}}\right),
\]
 
\[
R_{-2}=k_{-2}\left(1-\frac{k_{-2}+k_{2}P_{CO}\pm\sqrt{B_{2}}}{4k_{1}P_{O_{2}}}\right),
\]
 
\[
R_{3}=2R_{1}=2k_{1}P_{O_{2}}\left(\frac{k_{-2}+k_{2}P_{CO}\pm\sqrt{B_{2}}}{4k_{1}P_{O_{2}}}\right)^{2}.
\]
The equation $2R_{1}=R_{2}-R_{-2}$ is also valid for the second branch of the slow curve.

\textit{Bifurcation points.}

Now we will calculate the steady-state reaction rates at the bifurcation
points.

At the first bifurcation point $B_{1}=0$ (extinction point) the reaction
rates for the branch $y=0$ are: 
\[
R_{1}=k_{1}P_{O_{2}},
\]
 
\[
R_{2}=2k_{1}P_{O_{2}},
\]
 
\[
R_{-2}=0,
\]
 
\[
R_{3}=2R_{1}=2k_{1}P_{O_{2}}.
\]

After this bifurcation the system "jumps" to the stable steady state
on the branch $x=0$. The condition $B_1=0$ is fulfilled at the reaction rate $2k_{1}P_{O_{2}}\left(\frac{k_{-2}+k_{2}P_{CO}-\sqrt{B_{2}}}{4k_{1}P_{O_{2}}}\right)^{2}$ after the extinction point.
Thus, the main reaction rate of $R_{3}$ at this steady state
is

\[
R_{3,1}=\frac{(k_{-2})^{2}}{2k_{1}P_{O_{2}}}.
\]

 Because $R_{3}$ "jumps" from the larger value $2k_{1}P_{O_{2}}$ to the
smaller value $\frac{(k_{-2})^{2}}{2k_{1}P_{O_{2}}}$ this bifurcation
point is {\it the extinction point}, i.e it "jumps" from the extinction value $R_{A}$ to the "after extinction" value $R_{B}$. By previous calculations 

\begin{equation}
R_{A} R_{B}=(k_{-2})^{2}.\label{eq:firstproduct}
\end{equation}

At the second bifurcation point $B_{2}=0$ (i.e. $\left(k_{2}P_{CO}+k_{-2}\right)^{2}=8k_{-2}k_{1}P_{O_{2}}$)  reaction
rates for the branch $x=0$ are: 
\[
R_{1}=\frac{k_{-2}}{2},
\]
 
\[
R_{2}=\frac{k_{2}P_{CO}}{4k_{1}P_{O_{2}}}\left(k_{-2}+k_{2}P_{CO2}\right)=\frac{k_{2}P_{CO}}{4k_{1}P_{O_{2}}}\sqrt{8k_1P_{O_2}k_{-2}},
\]
 
\[
R_{-2}=\frac{k_{-2}}{4k_{1}P_{O_{2}}}\left(4k_{1}P_{O_{2}}-(k_{-2}+k_{2}P_{CO})\right)=\frac{k_{-2}}{4k_{1}P_{O_{2}}}\left(4k_{1}P_{O_{2}}-\sqrt{8k_1P_{O_2}k_{-2}}\right),
\]
 
\[
R_{3}=2R_{1}=k_{-2}.
\]

After this bifurcation the system "jumps" to the stable steady state
on the branch $y=0$. The main reaction rate $R_{3}$ at this steady state
is

\[
R_{3,1}=\frac{k_{2}^{2}P_{CO}^{2}}{2k_{1}P_{O_{2}}}.
\]
 
 Because $R_{3}$ "jumps" from the smaller value $k_{-2}$ to the bigger
value $\frac{k_{2}^{2}P_{CO}^{2}}{2k_{1}P_{O_{2}}}$ this bifurcation
point is {\it the ignition point}, i.e it "jumps" from the ignition value $R_{C}$ to the after ignition value $R_{D}$. By previous calculations 

\begin{equation}
R_{C} R_{D}=\frac{k_{-2}k_{2}^{2}P_{CO}^{2}}{2k_{1}P_{O_{2}}}.\label{eq:second product}
\end{equation}

At the bifurcation point of the maximal complexity $B_{1}=B_{2}=0$ we
have a further simplification: 
\[
R_{3}=k_{-2}=k_{2}P_{CO}=2k_{1}P_{O_{2}}.
\]

\vspace*{1cm}
\setcounter{equation}{0}
\section{Dynamics of Langmuir system}

In all figures, straight lines represent the zero approximation $\varepsilon=0$,
dashed lines represent the first approximation, thick points are steady
states, and small arrows represent system trajectories. All figures
are produced by the computer program ``Mathematica".

By standard stability analysis we have that for $B_{1}>0$
and $B_{2}>0$ the steady states of $(1,0)$ and 
\[
\left(0,1-\frac{k_{-2}+k_{2}P_{CO}+\sqrt{B_{2}}}{4k_{1}P_{O_{2}}}\right)
\]
 are saddle points. The steady states of $\left(1-\frac{k_{2}P_{CO}}{2k_{1}P_{O_{2}}},0\right)$
and 
\[
\left(0,1-\frac{k_{-2}+k_{2}P_{CO}-\sqrt{B_{2}}}{4k_{1}P_{O_{2}}}\right)
\]
 are stable (attractive) nodes.

If the initial conditions satisfy the following inequality: 
\[
x<y-1+\frac{k_{-2}+k_{2}P_{CO}+\sqrt{B_{2}}}{4k_{1}P_{O_{2}}}
\]
 all trajectories are  attracted to the node 
\[
\left(0,1-\frac{k_{-2}+k_{2}P_{CO}-\sqrt{B_{2}}}{4k_{1}P_{O_{2}}}\right).
\]

For any other initial conditions all trajectories are  attracted to the node
\[
\left(1-\frac{k_{2}P_{CO}}{2k_{1}P_{O_{2}}},0\right)
\]
 (see Figure 1). 
\begin{figure}[ht]
 ~\label{r1}

\centering{}\includegraphics[width=8cm]{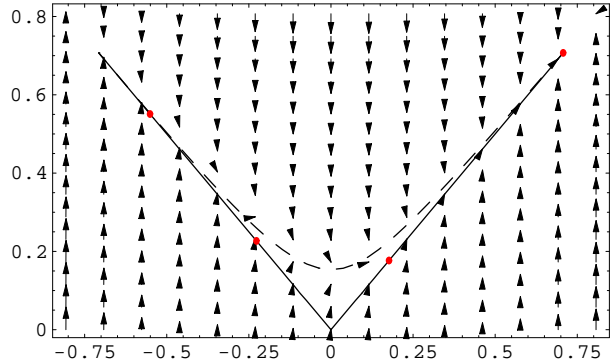} \protect\caption{Dynamics of the system in the case $B_{1}>0$, $B_{2}>0$. Solid line: zero approximation of slow manifold, dotted line: first approximation of slow manifold, red points: steady states.}
\end{figure}

In the case $B_{1}>0$, $B_{2}<0$ there are two steady states on
the branch $y=0$ and no steady state on the branch $x=0$. Hence,
all trajectories are  attracted to the node 
\[
\left(1-\frac{k_{2}P_{CO}}{2k_{1}P_{O_{2}}},0\right).
\]
 (see Figure 2).

\begin{figure}[ht]
 ~\label{r3}

\centering{}\includegraphics[width=8cm]{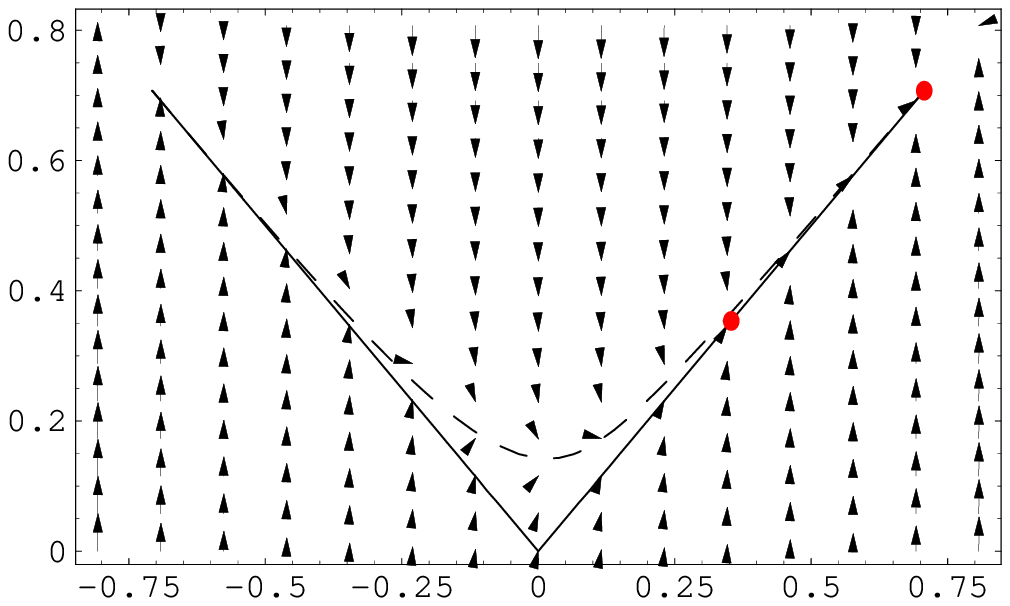} \protect\caption{Dynamics of the system in the case $B_{1}>0$, $B_{2}<0$. Solid line: zero approximation of slow manifold, dotted line: first approximation of slow manifold, red points: steady states.}
\end{figure}

In the case $B_{1}<0$, $B_{2}>0$ the steady state is  
\[
\left(0,1-\frac{k_{-2}+k_{2}P_{CO}+\sqrt{B_{2}}}{4k_{1}P_{O_{2}}}\right), 
\]
  and it does not belong to the invariant domain $U$. Hence we have only two singular
points. 

The first one $(1,0)$ is a saddle point, while  the second one is a stable
node 
\[
\left(0,1-\frac{k_{-2}+k_{2}P_{CO}-\sqrt{B_{2}}}{4k_{1}P_{O_{2}}}\right). 
\]
 All trajectories are attracted to the second point (see Figure 3).

\begin{figure}[ht]
 ~\label{r4a}

\centering{}\includegraphics[width=8cm]{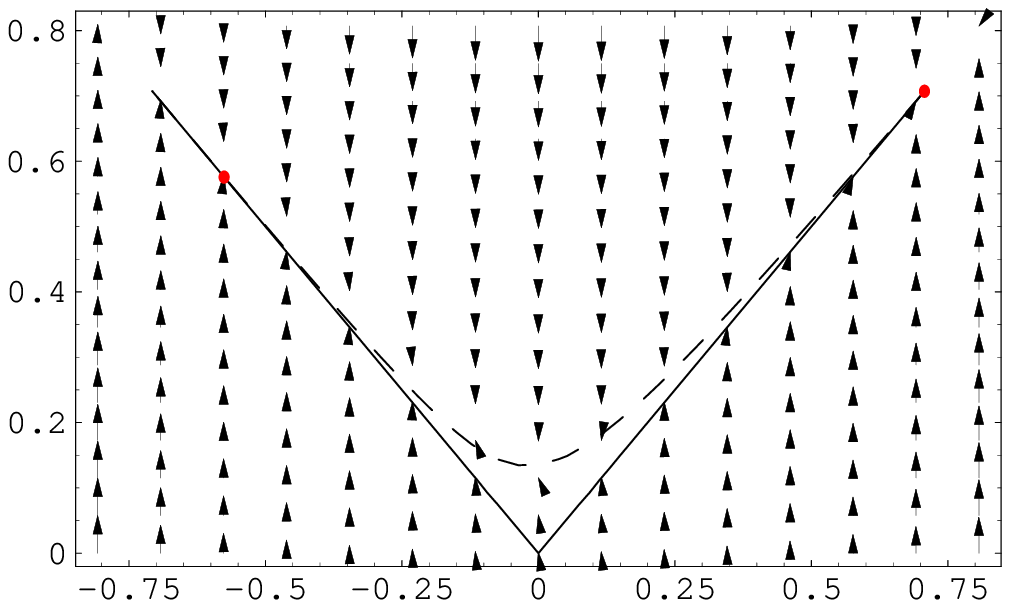} \protect\caption{Dynamics of the system in the case $B_{1}<0$, $B_{2}>0$. Solid line: zero approximation of slow manifold, dotted line: first approximation of slow manifold, red points: steady states.}
\end{figure}

In bifurcation cases where one or both parameters $B_{1}$, $B_{2}$
become zero more accurate approximations are necessary and such analyses
are beyond the scope of the zero approximation.

\textbf{Conclusion.} If $B_2>0$ and the sign of  the bifurcation
parameter $B_{1}$ is changed from the positive to negative one, the reaction
rate $R_{3}$ decreases dramatically from large to small values (compare figures 1 and 3).
It means that $B_{1}=0$ corresponds to the extinction point.

If $B_1>0$ and the sign of the bifurcation parameter $B_{2}$ is changed from the positive to negative one, the reaction rate $R_{3}$ make a "jump" from
small to larg values (compare figures 2 and 3). This means that
$B_{2}=0$ corresponds to the ignition point.

It is shown that  the point of maximal bifurcation complexity present a coincidence  of ignition and extinction points."

\vspace*{1cm}
\setcounter{equation}{0}
\section{About the phisyco-chemical meaning of bifurcation parameters and
bifurcations conditions}

We follow chapter 9 of Marin and Yablonsky's book \cite{YablBook}.

The condition $B_{1}=0$ is equivalent to $z=\frac{k_{2}P_{CO}}{2k_{1}P_{O_{2}}}=1$
for the corresponding equilibrium, i.e.  the entire catalyst surface is
empty at the equilibrium. Therefore, the bifurcation parameter $B_{1}:=1-\frac{k_{2}P_{CO}}{2k_{1}P_{O_{2}}}$
determines the difference between the complete concentration of catalytic
sites, i.e,unity, and the corresponding steady-state concentration of the empty catalytic sites. 

The condition $B_{2}=0$ can be reformulated in terms of the parameter

\begin{equation}
H:=\frac{\left(k_{2}P_{CO}+k_{-2}\right)^{2}}{8k_{1}P_{O_{2}}k_{-2}}.\label{eq:H}
\end{equation}
 The condition $B_{2}=0$ is equivalent to $H=1$. 

By the previous analysis the condition $B_{1}=0$ determines the extinction
point and the condition $B_{2}=0$ or $H=1$ determines the ignition
point. 

Detailed explanations of the condition $H=1$ can be found in the
book (\cite{YablBook} , chapter 9). We repeat here these arguments.
``At the ignition point, the steady state reaction rate $R_{3}=k_{3}xy$
is determined only by the $CO$ desorption rate coefficient. It does not
depend on the composition of the gas-phase mixture.'' At the ignition
point the concentration $z_{ig}$ of active sites is equivalent to
$2z_{eq}$, where $z_{eq}$ is an equilibrium concentration of free
active sites. Let us discuss in more detail the concept of equilibrium concentration. Suppose that in our system we have only one reversible adsorption step, i.e. an interaction between the empty catalytic sites and gaseous $CO$. Then at the equilibrium which in this case is the 
steady state as well 
\[
k_{2}P_{co}z_{eq}=k_{-2}y_{eq}.
\]

From these equations, obviously

\[
z_{eq}=\frac{k_{-2}}{k_{2}P_{CO}+k_{-2}}=\frac{1}{K_{eq,2}P_{CO}+1},
\]
where $K_{eq,2}$ is the equilibrium constant
of $CO$ adsorption.

Before ignition, the steady state concentration of adsorbed
oxygen, $x$ is very low,  therefore in the vicinity of ignition we approximately have 
\[
y=1-z_{ign}=1-2z.
\]

At the point of the maximal bifurcational complexity both conditions $B_1=0$ and $H=1$ are fulfilled. It means that at this point {\bf two phenomena, ignition and extinction, merge.}
All the catalyst sites are empty. Using $z=2z_{eq}=1$, we have
\[
z_{eq}=\frac{2}{K_{eq,2}P_{CO}+1}=1
\]
 
and $K_{eq,2}P_{CO}=1$.

The same equation can be obtained in a more similar way, because at the point of the maximal bifurcational complexity (MBC-point) ignition and extinction are merged, and the following equation is valid
\[
R_{ext}=R_{ign}
\] 
where $R_{ext}$ and $R_{ign}$ are reaction rates at ignition and extinction points respectively.
Consequently $k_{-2}=k_2 P_{CO}$ and $K_{eq,2}P_{CO}=1$. 

The last equation is quite interesting. It is nothing but the 'degenerate'   
Langmuir equation in the limit case when all the catalytic surface is    
covered by $CO$ (rather to say almost covered).

This equation gives a unique possibility to estimate Keq based just on   
the  $MBC$ -point experimentally observed:
\[
K_{eq,2}=1/P_{CO,MBC},
\]

where $P_{CO,MBC}$ is the partial pressure of $CO$ at the $MBC$-point.

\vspace{0.5cm}

{\bf Remark} Distinguishing the well-observed and ill observed critical characteristics and parameters.

 \vspace{0.5cm}

In catalytic $CO$ oxidation over the $Pt$ catalyst,  the reaction rates at ignition and extinction points  are ill -observed, but the after-ignition and after-extinction points are well-observed.  It is possible to find reaction rates of the ill-defined points (ignition and extinction) and parameters of adsorption steps based on the information regarding the well-defined points (after-ignition and after-extinction).

\setcounter{equation}{0}
\section{Conclusions}

A phenomenon of the bifurcational parametric simplification was presented as a generalization and a justification of the critical simplification principle described in \cite{SimpleModel}. The parametric simplification was applied to the Langmuir model using its transformation to the slow-fast dynamic system based on the concept of singularly perturbed vector fields \cite{BGG2006}. A detailed analysis of the Langmuir model was performed, and a bifurcation of the maximum complexity was presented as a new peculiarity in terms of kinetic parameters.

\vspace*{1cm}
\setcounter{equation}{0}
\section*{Appendix 1. Mathematical models and the structure of chemical reaction
mechanisms}

Additional definitions of the chemical source term are required
for the exposition. A pure homogeneous system is considered. It is
represented by a system of ordinary differential equations (ODEs)
that describes the mechanism of chemical kinetics by a system based
on the mass action law.

The mathematical model describes the temporal evolution of chemical state vector,
where $\psi_j$ represents a concentration of the its chemical substance. 
A dimension of the system is the number of species (reactants), $n$: 
\[
\psi=(\psi_{1},...,\psi_{n})^{T}.
\]

In vector notation the system of governing equations of a homogeneous
system can be written in autonomous form as (see e.g. \cite{YBE1984,W1985,L1997})

\begin{equation}
\frac{d\psi}{dt}=F(\psi),\quad\psi\in\Omega\subset R^{n}.\label{eq:1}
\end{equation}

Here the so called chemical source $F$ represents the chemical mechanism:  $n_{s}$ reactants participate in $n_{r}$ elementary chemical reactions.  Pressure and enthalpy are considered to be constant. 

The elementary reactions are 

\begin{equation}
i=1,...,n_{r}:\:\alpha_{1,i}A_{1}+...+\alpha_{n_{s},i}A_{n_{s}}\leftrightarrow\beta_{1,i}A_{1}+...+\beta_{n_{s},i}A_{n_{s}}\label{eq:2}
\end{equation}
where $A_{1,...,}A_{n_{s}}$ chemical species
which participate in $n_r$ elementary reactions.
Reaction rates of these reactions relate to the mass action law which was already mentioned [7]. This law implies the polynomial form 
\begin{equation}
R_{i}(\psi):=k_{i}^{+}\prod_{j=1}^{n_{s}}\psi_{j}^{\alpha_{j,i}}-k_{i}^{-}\prod_{j=1}^{n_{s}}\psi_{j}^{\beta_{j,i}}\quad i=1,...,n_{r}.\label{eq:3}
\end{equation}
The rate constants $k_{i}^{+},\, k_{i}^{-}$ are characterized by exponential Arrhenius dependences on the temperature.

Therefore, components of $F(\psi)$ are composed in the following
way 
\begin{equation}
F_{j}(\psi)=\sum_{i=1}^{n_{r}}\gamma_{j,i}R_{i}(\psi), \label{eq:4}
\end{equation}
 where $\gamma_{j,i}:=\beta_{j,i}-\alpha_{j,i}$ are components of
the so-called stoichiometric coefficient ($n_{s}\times n_{r}$) matrix
$S$, and $R(\psi)$ is a non-linear vector function of the elementary
reaction rates.

A specific structure of models of chemical kinetics is discussed now.
The structure is based on the stoichiometric matrix $S$. It is typical
in chemical kinetics that the number of reactions is much larger then
the number of species i.e. $n_{r}\gg n_{s}$ and $S$ is a rectangular
matrix of a linear mapping from the reaction space $R^{n_{r}}$ to
the system state space $R^{n_{s}}$. The vector of elementary reaction
rates $R(\Psi)$ is a non-linear map from the state space $R^{n_{s}}$
to the reaction space $R^{n_{r}}$. Their composition represents the
chemical source term (vector) $F(\Psi)$ that maps the state space
$R^{n_{s}}$ into itself.

This special structure of kinetic models (\ref{eq:1}) - (\ref{eq:4})
can be illustrated by the following diagram

\[
\begin{array}{ccccc}
 &  & R^{n_{r}}\\
 & R(\Psi)\\
 & \nearrow & \downarrow & S\\
R^{n_{s}} & \longrightarrow & R^{n_{s}}\\
 & F(\Psi)
\end{array}
\]

The initial linear reduction procedure is possible for the linear
map represented by the matrix $S$. Denote by 
\[
\ker S=\{w\in R^{n_{r}}|S(w)=0\}
\]
 the kernel of the linear map $S:\, R^{n_{r}}\to R^{n_{s}}$. It is
clear that $\dim(\ker S)\geq n_{r}-n_{s}$. If $\dim(\ker S)>n_{r}-n_{s}$,
then there exist 
\[
q=\dim(\ker S)-\left(n_{r}-n_{s}\right)>0
\]
 linear integrals in the original system. Hence, without loss of generality,
the first $n_{s}-q$ lines of $S:=\left\Vert \gamma_{j,i}\right\Vert $
can be assumed to be linearly independent, while the $q$ lines $k=n_{s}-q+1,...,\, n_{s}$
are linearly dependent 
\[
\gamma_{k,\cdot}=\sum_{\alpha=1}^{n_{s}-q}\sigma_{k,\alpha}\gamma_{\alpha,\cdot}.
\]
 As the result, the following linear integrals of the original system
can be found, given by the $\left(q\times n_{s}\right)$ matrix

\[
\tilde{Z_{c}}=\left(\begin{array}{ccccccc}
\sigma_{1,1} & ... & \sigma_{1,n_{s}-q} & -1 & 0 & ... & 0\\
\sigma_{2,1} & ... & \sigma_{2,n_{s}-q} & 0 & -1 & ... & 0\\
... & ... & ... & ... & ... & ... & ...\\
\sigma_{q,1} & ... & \sigma_{q,n_{s}-q} & 0 & 0 & ... & -1
\end{array}\right),
\]

\[
\tilde{Z_{c}}\cdot S\equiv0,
\]
 which means $\tilde{Z_{c}}$ defines the conserved quantities of
the system

\[
\tilde{Z_{c}}\cdot\frac{d\psi}{dt}=\tilde{Z_{c}}\cdot F(\psi)=\tilde{Z_{c}}\cdot S\cdot R(\psi)=0,
\]
 i.e. 
\[
\tilde{Z_{c}}\cdot\left(\psi-\psi^{0}\right)=0.
\]
 This procedure permits us to reduce the dimension of the system using the linear integrals which typically have a simple physic-chemical meaning corresponding to conserved quantities of chemical elements.

A detailed study of $\ker S$ has a long history. It is known as the
``methabolic pathway analysis of null space of stochiometric matrix''
(see, for example, \cite{Clarke}). A more detailed analysis of linear-algebraic
approaches  with help of molecular and Horiuti matrices can be found
in \cite{ConstYablMar}.

Steady states $\Psi_{1},...\Psi_{m}$ of this model are solutions
of the following analytic equation 
\[
F(\Psi_{i})=0,i=1,2,...,m.
\]
 Its coordinates are functions of main system parameters. Therefore,
the linear integrals produce functional dependencies between main
model parameters. Solving these functional dependencies we reduce
number of main independent parameters by the number of the linear integrals.
Typically explicit solutions of the steady state equations and its
consequences for main system parameters, have a very complicated analytic
nature because of the high non-linearity of the original models.

It is a first step of the parametric critical simplification.

The second step of the parametric critical simplification is connected
with bifurcations. The notion of a bifurcation is not completely formal.
In the case of steady state multiplicity, it means coincidence of two
different steady states i.e. $\Psi_{i}=\Psi_{j},i\neq j$. It gives an additional functional dependence between the main parameters of the model and therefore permits us to reduce at least one independent parameter. Because bifurcation typically has a simple physical
meaning (ignition or extinction points, appearance of oscillations
etc.) it permits us to check values of leading parameters experimentally.

\vspace*{1cm}
\setcounter{equation}{0}
\section*{Appendix 2. About the first approximation of the slow invariant curve}

Recall that in $u,v$-coordinates the Langmuir system has
the form 
\[
\frac{du}{dt}=\sqrt{2}k_{1}P_{O_{2}}(1-\sqrt{2}v)-\frac{k_{2}}{\sqrt{2}}P_{CO}(1-\sqrt{2}v)^{2},
\]
 
\[
\varepsilon\frac{dv}{dt}=\frac{u^{2}-v^{2}}{\sqrt{2}}+\frac{\varepsilon}{2\sqrt{2}}\left(4k_{1}P_{O_{2}}(1-\sqrt{2}u)^{2}+2k_{2}P_{CO}(1-\sqrt{2}v)\right).
\]

Because the zero approximation $\varepsilon=0$ is not informative
at the intersection point $(0,0)$ of the two branches $u=v$ and $u=-v$,
we calculate the first approximation of the slow curve. It is 
\[
u^{2}-v^{2}+\varepsilon\left(k_{2}P_{CO}(1-\sqrt{2}v)^{2}\frac{u+v}{v}+2k_{1}P_{O_{2}}(1-\sqrt{2})^{2}\frac{v-u}{v}\right)=0
\]
 (see Figure 4).

We give here a short sketch of the first approximation of steady
state multiplicity. We do not present here all of the long elementary
calculations, but focus rather on the vicinity of the problematic intersection
point $(0,0)$. Far from $(0,0)$ the first approximation loses its corrective influence on the results.

The first bifurcation parameter $B_{1}$ in the first approximation
becomes 
\[
B_{1,1}=(1-\frac{k_{2}P_{CO}}{2k_{1}P_{O_{2}}})^{2}-\varepsilon\frac{2k_{2}^{2}P_{CO}^{2}}{k_{1}P_{O_{2}}}.
\]
 Depending on this bifurcation parameter the system has four, three
or two steady states. Existence of three steady states ($(0,1)$,
$(1,0)$ and 
\[
(\frac{1}{2}\sqrt{\varepsilon\frac{2k_{2}^{2}P_{CO}^{2}}{k_{1}P_{O_{2}}}},\frac{1}{2}\sqrt{\varepsilon\frac{2k_{2}^{2}P_{CO}^{2}}{k_{1}P_{O_{2}}}})
\]
 corresponds to $B_{1,1}=0$. The third steady state 
\[
(\frac{1}{2}\sqrt{\varepsilon\frac{2k_{2}^{2}P_{CO}^{2}}{k_{1}P_{O_{2}}}},\frac{1}{2}\sqrt{\varepsilon\frac{2k_{2}^{2}P_{CO}^{2}}{k_{1}P_{O_{2}}}})
\]
 belongs to an $\varepsilon$-vicinity of $(0,0)$. For $\varepsilon\rightarrow0$
this point is disappears from the vicinity of the (0,0) point. This
steady state  disappears for $B_{1,1}<0$.

 The first order approximation of the slow manifold is presented at Figure 4.

\begin{figure}[ht]
~\label{r4}

\centering{}\includegraphics[width=8cm]{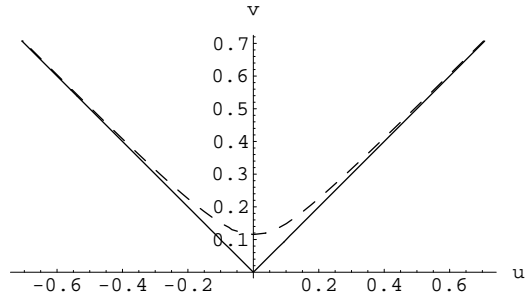} \protect\caption{Solid line: zero approximation of slow manifold, dotted line: first
approximation of slow manifold }
\end{figure}

In the invariant domain we have that $k_{2}P_{CO}(1-\sqrt{2}v)^{2}\frac{u+v}{v}+2k_{1}P_{O_{2}}(1-\sqrt{2})^{2}\frac{v-u}{v}>0$.
This inequality leads to the conclusion that non trivial steady states
lie inside the invariant domain. In the case $B_{1,1}=0$ two non trivial
steady states come together and we obtain one steady state that it
is a saddle-node. In this case, if the initial condition satisfies the
inequality: $x<y$, all trajectories attract to the $(0,1)$ point.
If the initial condition satisfies the inequality $x>y$, all trajectories
attract to the $\left(\frac{1}{2}\sqrt{\varepsilon\frac{2k_{2}^{2}P_{CO}^{2}}{k_{1}P_{O_{2}}}},\frac{1}{2}\sqrt{\varepsilon\frac{2k_{2}^{2}P_{CO}^{2}}{k_{1}P_{O_{2}}}}\right)$
point.

The case of $B_{1}=0$, $B_{2}=0$ is more complex and the standard asymptotic
analysis is not applicable. Actually, this case is more interesting
because this bifurcation point is the point of the maximal simplification
of the system. At this point we obtain that $k_{-2}=k_{2}P_{CO}=2k_{1}P_{O_{2}}$. 

\vspace{1cm}

{\bf Acknowledgment}

\vspace{0.5cm}

The financial support by GIF (Grant 1162-148.6/2011) project is gratefully acknowledged.

We thank Professor Alexander Gorban for discussions and comments that greatly improved the paper.

\end{document}